\documentclass[floatfix,twocolumn,aps,prd,showpacs,amsmath,amssymb]{revtex4}

\selectfont 

\newcommand{\alp}{\alpha}

\newcommand{\beq}{\begin{equation}}
\newcommand{\eeq}{\end{equation}}

\newcommand{\RR}{\chi}

\def\b{\begin{equation}}
\def\e{\end{equation}}
\def\balll{\begin{array}{lll}}
\def\ea{\end{array}}
\def\bea{\begin{eqnarray}}
\def\eea{\end{eqnarray}}

\usepackage{epsfig}
\usepackage{color}

\begin{document}
 \title{Quasinormal Modes and Regge Poles of the Canonical Acoustic Hole}

\author{Sam R. Dolan}
\email{s.dolan@soton.ac.uk}
\affiliation{School of Mathematics, University of Southampton,
Highfield, Southampton SO17 1BJ, United Kingdom}

\author{Leandro A. Oliveira}
\email{laoliveira@ufpa.br}
\affiliation{Faculdade de
F\'\i sica, Universidade Federal do Par\'a, 66075-110, Bel\'em,
Par\'a, Brazil}

\author{Lu\'\i s C. B. Crispino}
\email{crispino@ufpa.br}
\affiliation{Faculdade de F\'\i sica, Universidade Federal do
Par\'a, 66075-110, Bel\'em, Par\'a, Brazil}

\date{\today}

\begin{abstract}
We compute the quasinormal mode frequencies and Regge poles
of the canonical acoustic hole (a black hole analogue), using three methods. First, we show how damped oscillations arise by evolving generic perturbations in the time domain using a simple finite-difference scheme. We use our results to estimate the fundamental QN frequencies of the low multipolar modes $l=1, 2, \ldots$. Next, we apply an asymptotic method to obtain an expansion for the frequency in inverse powers of $l+1/2$ for low overtones. We test the expansion by comparing against our time-domain results, and (existing) WKB results. The expansion method is then extended to locate the Regge poles. Finally, to check the expansion of Regge poles we compute the spectrum numerically by direct integration in the frequency domain. We give a geometric interpretation of our results and comment on experimental verification. 
\end{abstract}


\maketitle

%
%

\section{Introduction}

Black hole quasinormal (QN) modes are characteristic damped 
resonances which depend only on the parameters of the black hole
(mass, charge and angular momentum)~\cite{Kokkotas-Schmidt}. A slightly-perturbed black hole returns to equilibrium by shedding its asymmetries (i.e. higher multipoles) through gravitational radiation. The radiated signal carries the imprint of the least-damped quasinormal modes. In principle, a detection of QN frequencies in a gravitational wave signal would allow direct inference of black hole mass and angular momentum. Given this motivation it is no surprise that QN spectra have been extensively investigated along the years for black hole systems and relativistic stars (see, e. g., \cite{Nollert, Ferrari-Gualtieri, Berti-Cardoso-Starinets} and references therein). QN modes also play a role in the dynamics of other `open' systems in physics \cite{Ching1}, for example, radiation in an optical cavity \cite{Ching2, Severini}; photonic crystals \cite{photonic-crystals}; the AdS/CFT correspondence \cite{Nunez-Starinets}; and holographic QCD models \cite{Evans-Threlfall}. 

Of course it is not possible to study a black hole directly in the laboratory; but ever since Unruh's original proposal \cite{Unruh} various kinds of `black hole analogues' have been suggested. A black hole analogue is a non-gravitational system that mimics key features of a black hole, such as horizons and ergoregions. A range of analogue systems have been suggested, for example in acoustics \cite{Unruh, Visser, Visser-1993}, superfluid helium~\cite{Volovik-2003}, in Bose-Einstein condensates~\cite{Garay-2002}, in electromagnetic waveguides~\cite{Schutzhold-Unruh-2004}, in optical fibers~\cite{Corley-Jacobson-1999, Unruh-Schutzhold-2007} and other systems~\cite{Novello-Visser-Volovik}. Although analogue systems are not subject to Einstein equations, they do allow, in principle, a realization of all \emph{kinematic} aspects of general relativity. Hence such systems could enable an experimental study of the propagation of classical and quantum fields in curved spacetime \cite{Visser-Barcelo-Liberati}. For instance, the generation of analogue horizons may enable the experimental detection of an analogue form of Hawking radiation (see \cite{Weinfurtner-et-al} for recent progress). Further motivations for the study of analogue systems are described in \cite{Barcelo-Liberati-Visser}.

Acoustic analogues are motivated by a key observation \cite{Unruh,Visser-1993,Visser}: under certain conditions, small perturbations $\delta \mathbf{v} = -\nabla \Phi$ to a smooth fluid flow $\mathbf{v}$ are governed by  
\begin{equation}
\nabla_\nu \nabla^\nu \Phi=\frac{1}{\sqrt{-g}}
\partial_\mu \left(\sqrt{-g} g^{\mu \nu} \partial_\nu \Phi \right)=0 ,
\label{klein}
\end{equation}
where $g_{\mu \nu}$ is a function of the local properties of the fluid flow (see \cite{Unruh, Visser-1993, Visser} for details). Eq.~(\ref{klein}) is formally equivalent to the minimally-coupled Klein-Gordon equation on a background spacetime described by the metric $g_{\mu \nu}$. Hence it is natural to interpret $g_{\mu \nu}$ as describing an \emph{effective geometry} on which perturbations propagate. Note that $g_{\mu \nu}$ is not a solution of Einstein's equations; instead the effective geometry is set by the background flow, which is under experimental control. Naturally, the prospect of observing the propagation of perturbations on curved spacetimes in the laboratory has attracted widespread interest. 

An `idealized' example of an acoustic analogue system is the `canonical' acoustic hole (CAH) described in \cite{Visser}. The CAH is formed from a stationary spherically-symmetric flow in an inviscid barotropic fluid of constant density. Small perturbations to the CAH are governed by Eq.~(\ref{klein}) with an effective geometry $ds^2 = g_{\mu \nu} dx^{\mu} dx^{\nu} $ where \cite{time-footnote}
\beq
ds^2 = f (c dt)^2-f^{-1}dr^2-r^2\left(d\theta^2+\sin^2{\theta}d\varphi^2\right),   \label{metric}
\eeq
and 
\beq
f(r) = 1 - r_h^4 / r^4 \label{eq-f} .
\eeq
Here $r_h$ is the horizon radius, where the flow becomes supersonic (i.e. where the radial flow speed exceeds the speed of sound $c$). Henceforth we set the speed of sound equal to unity ($c=1$). 

Line element (\ref{metric}) bears a striking similarity to the well-known Schwarzschild solution, for which $f = 1 - r_h / r$. Just like the Schwarzschild solution, the CAH has a spectrum of QN modes, labeled by angular momentum $l$ and overtone number $n \ge 0$. 

The QN modes of the CAH were studied in \cite{Berti-Cardoso-Lemos}, 
 using the WKB method. The authors obtained estimates of the frequencies of the low overtones (see Table \ref{table1}). QN modes were also the subject of a recent time domain study \cite{Xi-Li}. Various authors have considered QN modes in more realistic laboratory configurations \cite{Cardoso-Lemos-Yoshida, Abdalla-Konoplya, Nakano-Kurita}. Precise determination of QN frequencies of the CAH remains an open challenge, in part because the method of Jaff\'e series and continued fractions \cite{Leaver-1985} (successfully applied to the black hole case) cannot be easily extended to treat the CAH due to the $1/r^4$ factor in $f$ in Eq.~(\ref{eq-f}). However, there exists a wide range of other methods for computing QN modes: see \cite{Berti-2004} or \cite{Dolan-Ottewill} for partial lists, and new methods \cite{Cho-Cornell, Denef-Hartnoll-Sachdev}. In this paper we make use of three approaches: time-domain simulation \cite{Davis-Ruffini, Press, Dorband, Okuzumi-Sakagami}, asymptotic expansion \cite{Dolan-Ottewill} and direct integration in the frequency domain  \cite{Chandrasekhar-Detweiler}. We demonstrate the connection between QN modes and the Regge poles of scattering theory \cite{newton} which were studied in a black hole context in \cite{Decanini-Folacci, Decanini-Folacci-Raffaelli}.  

The remainder of this paper is structured as follows: in Sec.~\ref{sec-basics} we cover the theory of perturbations of the CAH, define both quasinormal modes and Regge poles, and study the circular orbits of null geodesics on the effective spacetime; in Sec.~\ref{sec-timedomain} we evolve generic perturbations in the time domain and identify quasinormal mode ringing; in Sec.~\ref{sec-QNexpansion} we recap the expansion method of \cite{Dolan-Ottewill} and apply it to find the QN modes of the CAH; in Sec.~\ref{sec-RPexpansion} we extend the method to find Regge poles, and introduce a numerical method for checking the expansion; we conclude in Sec.~\ref{sec-conclusion} with a brief discussion.

\section{Basics\label{sec-basics}}

 \subsection{Perturbations}
 Small perturbations of the fluid flow $\delta \mathbf{v} = - \nabla \Phi$ are governed by the Klein-Gordon equation (\ref{klein}) with effective geometry (\ref{metric}). Let us take advantage of spherical symmetry to decompose solutions of Eq.~(\ref{klein}) into
\begin{equation}
\Phi = \sum_{lm} \Phi_{lm}, \quad \Phi_{lm}= r^{-1} \psi_{l}\left(t, r\right) 
Y_{l m}\left(\theta,\phi\right).
\label{Phi}
\end{equation}
A modal perturbation $\psi_l(t,r)$ evolves according to the homogeneous wave equation
\beq
\left( \frac{ \partial^2 }{\partial t^2} - \frac{\partial^2 }{\partial r_\ast^2} + V_l(r) \right)  \psi_l = 0 , \label{cah-pde}
\eeq
with effective potential
\beq
V_l(r) = f(r) \left( \frac{l(l+1)}{r^2} + \frac{f^\prime(r)}{r}  \right) ,
\eeq
and `tortoise coordinate' $r_\ast$ defined by $ \frac{dr_\ast}{dr} = f^{-1} $,
\beq
r_*=r+\frac{r_h}{4}\ln \left|\frac{r-r_h}{r+r_h}\right|-\frac{r_h}{2} 
\arctan\left(\frac{r}{r_h}\right) +r_h \frac{\pi}{4}. \label{rstar} \\
\eeq
Physical perturbations are `ingoing' at the horizon 
\beq
\lim_{r_\ast \rightarrow -\infty} \left[ \frac{\partial \psi_l}{\partial t} - \frac{\partial \psi_l}{\partial r_\ast} \right] = 0 .
\eeq
Further decomposition via an integral transform 
\beq
\psi_l(t,r) = \frac{1}{2 \pi} \int_{-\infty + i \kappa}^{\infty + i \kappa} e^{-i \omega t} \varphi_{\omega l} (r) d\omega
\eeq
leads to the ordinary differential equation 
\beq
\left[ \frac{d^2}{d r_\ast^2} + \omega^2 - V_l(r)  \right] \varphi_{\omega l} = 0  \label{cah-rad-eq} 
\eeq
for single-frequency modes $\varphi_{\omega l}(r)$.

\subsection{Quasinormal modes and Regge poles\label{sec-QNRP}}
Quasinormal (QN) modes and Regge poles (RP) are special single-frequency modes which 
are purely ingoing at the horizon, and purely outgoing at spatial infinity~\cite{Kokkotas-Schmidt, Berti-Cardoso-Starinets}
\beq
\varphi_{l n}(r)=
\left\{
\begin{array}{ll}
e^{-i\omega r_*}, \hspace{0.5 cm} r_*\rightarrow -\infty, \\
A^{\text{(out)}}_{l n}e^{i\omega r_*}, \hspace{0.5 cm} r_*\rightarrow \infty .
\end{array}
\right.
\label{cah-bc}
\eeq
There exists a discrete spectrum of QN modes $\varphi_{l n}(r)$ and frequencies $\omega_{ln}$, labeled by angular multipole $l = 0, 1, \ldots$ and overtone number $n = 0, 1, \ldots$. QN frequencies are complex: the real part determines the oscillation frequency, and the (negative) imaginary part determines the decay rate. 

Regge poles are modes $\varphi_{\omega n}$ that obey boundary conditions (\ref{cah-bc}), with a \emph{real} frequency and a \emph{complex} angular momentum $\lambda_{\omega n} = l_{\omega n} + 1/2$. More precisely, they are poles of the S-matrix lying in the first quadrant of the complex angular momentum plane \cite{Decanini-Folacci}.  They are a key concept in, e.g., high-energy physics, where they are used to describe diffraction features \cite{newton}. 

 \subsection{Geodesics\label{sec-geodesics}}
High-frequency perturbations propagate along the null geodesics of the effective geometry. Geodesics were considered in detail in a previous study of scattering by the CAH \cite{doc}. We recall the key results below. 

The paths of null geodesics are obtained from the orbital equation 
\begin{equation}
\left(\frac{d u}{d \phi}\right)^2 = \frac{1}{b^2} - u^2 + r_h^4 u^{6} , \label{orbit-eq-1}
\end{equation}
where $u = 1/r$, and $b$ is the impact parameter. 
By solving Eq.~(\ref{orbit-eq-1}) we may obtain the deflection angle as a function of the impact parameter, $\Theta(b)$, 
\begin{equation}
\Theta(b) = \frac{2 K(k)}{\sqrt{(v_2 - v_1) v_3}}  - \pi, 
\quad \text{where} \quad 
k^2 = \frac{v_2 (v_3 - v_1)}{v_3 (v_2 - v_1)}.  \nonumber
\end{equation}
Here, $v_1$, $v_2$ and $v_3$ are the roots of the cubic
\begin{equation}
r_h^4 v^3 - v + 1/b^2 = 0 .  \label{cubic-roots}
\end{equation}
In the `critical' case $b = b_c$, Eq.~(\ref{cubic-roots}) has a repeated root at $r = r_c$, where
\beq
r_c = 3^{1/4} r_h , \quad \quad b_c  = r_c / \sqrt{f(r_c)} = \left( 3^{3/4} / 2^{1/2} \right) r_h .   \label{circ-orb-params}
\eeq
Rays with impact parameters $b >b_c$ are scattered; rays with $b < b_c$ are absorbed; and in the critical case, the ray with $b=b_c$ ends in perpetual orbit at $r=r_c$. In the critical case, the orbital equation (\ref{orbit-eq-1}) can be factorized into
\beq
b_c^2 \left(\frac{d u}{d \phi}\right)^2 = \left(1 - r_c^2 u^2 \right)^2 \left(1 + \frac{1}{2} r_c^2 u^2  \right)  . \label{orbit-eq-1c}
\eeq
In the limit $l \gg n$, it turns out that the spectrum of QN modes and RPs is determined by the properties of the null orbit \cite{Goebel, Mashhoon, Hod, Lyapunov1}. 
The frequency $\Omega$ of the null orbit is
\beq
\Omega = \dot{\phi} / \dot{t} = b_c f_c / r_c^2 = 1 / b_c  \label{orbital-freq} ,
\eeq
where $f_c = f(r_c)$, and its Lyapunov exponent $\Lambda$ \cite{Lyapunov1, Lyapunov2, Lyapunov3} is
\beq
\Lambda = \left( - \frac{b_c^2}{2} \frac{d^2}{dr_\ast^2} \frac{f}{r^2} \right)^{1/2}_{r=r_c} = 2 / b_c  \label{lyapunov-exp}
\eeq
(see, e.g., Eq. (40) in \cite{Lyapunov1}).

\section{Time-Domain Evolution\label{sec-timedomain}}
To observe the role that QN modes play in time-dependent scattering, we studied the evolution of a small perturbation of the CAH.  We imposed Gaussian initial data on the surface $t=0$, i.e. the initial condition
\begin{eqnarray}
\psi_l(t=0, r) &=& \exp\left(-(r_\ast - \bar{r}_\ast)^2 / (2 \sigma^2)\right),  \nonumber \\
\frac{\partial \psi_l }{\partial t} (t=0, r) &=& 0 ,
\end{eqnarray}
with some midpoint $\bar{r}_\ast$ and width $\sigma$. We simulated the evolution of this perturbation using a 1+1D finite difference scheme \cite{nr}.

\begin{figure}
\begin{center}
\includegraphics[width=8cm]{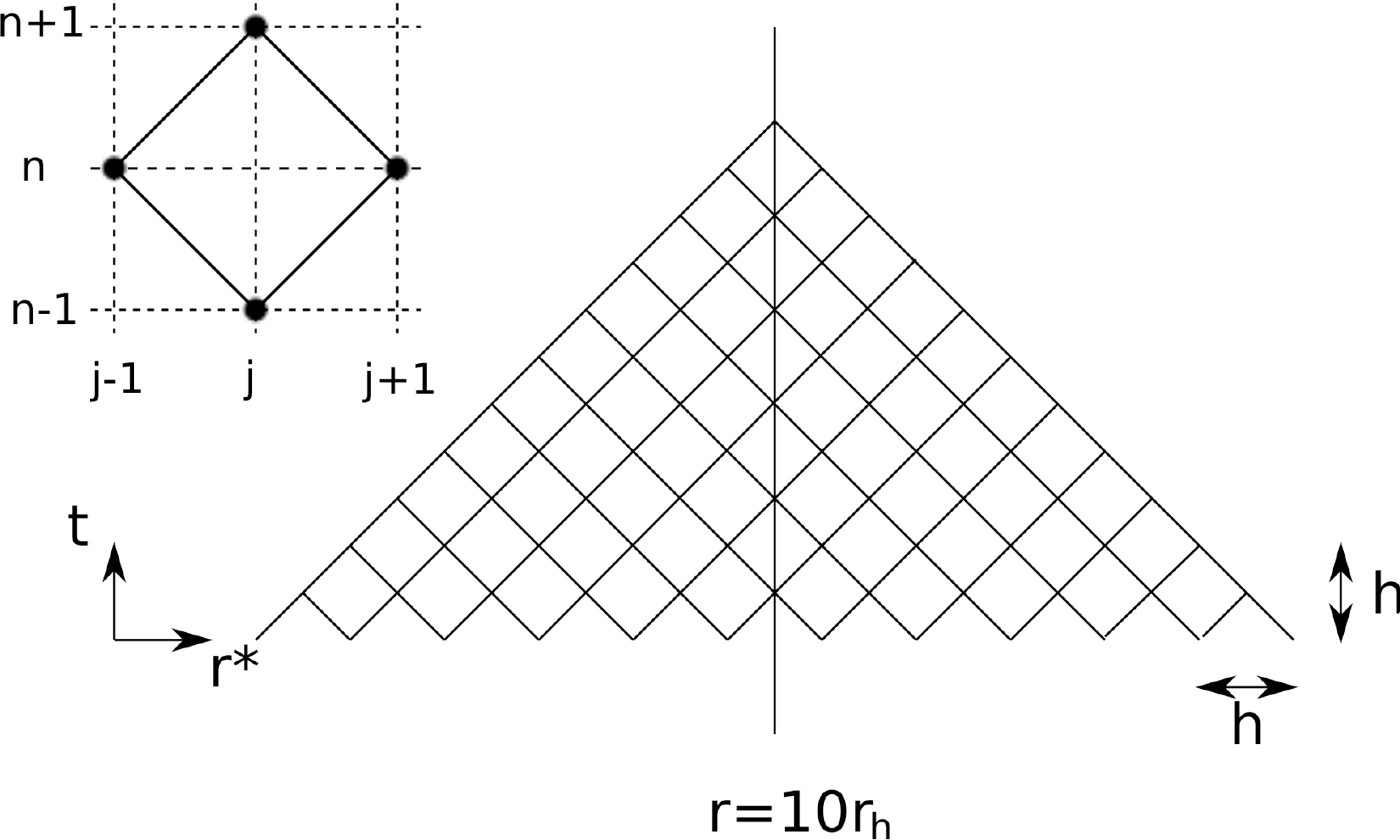}
\end{center}
\caption{\emph{Illustration of finite difference scheme}. The diagram shows a double-null grid, with spacing $\delta r_\ast = \delta t = h$. The field shown in Fig.~\ref{fig-timedomain} was extracted at $r = 10 r_h$. }
\label{fig-finitediff}
\end{figure}

The method is illustrated in Fig.~\ref{fig-finitediff}. We evolve on a grid in $(r_{\ast}, t)$ of resolution $\delta r_\ast = \delta t = h$ with $h = r_h/64$ for $t = 100 r_h$. The grid is made sufficiently large that the boundaries do not come into causal contact with the region of interest. We employ the scheme,
\beq
\psi_{j}^{n+1} =  \psi_{j+1}^{n} + \psi_{j-1}^n - \psi_j^{n-1} - \frac{h^2 V_j}{8} \left(  \psi_{j+1}^n + \psi_{j-1}^n \right)
\eeq
where $\psi_j^n \approx \psi_l(t_n, r^\ast_j)$, $V_j = V_l(r_j)$ and $r^\ast_j = j h / 2$, $t_n = n h / 2$. 
A similar scheme was employed in \cite{Gundlach-Price-Pullin, Barack-Golbourn}, in which it was shown that the global accumulated discretization error scales with $h^2$. The method is straightforward to implement, and the only numerical difficulty arises in inverting (\ref{rstar}) to obtain $r(r_\ast)$. We applied a numerical root finder (for $r_\ast > 0$) and an iterative method based on series expansions around $r=r_h$ (for $r_\ast < 0$).  

\begin{figure*}
\begin{center}
\includegraphics[width=16cm]{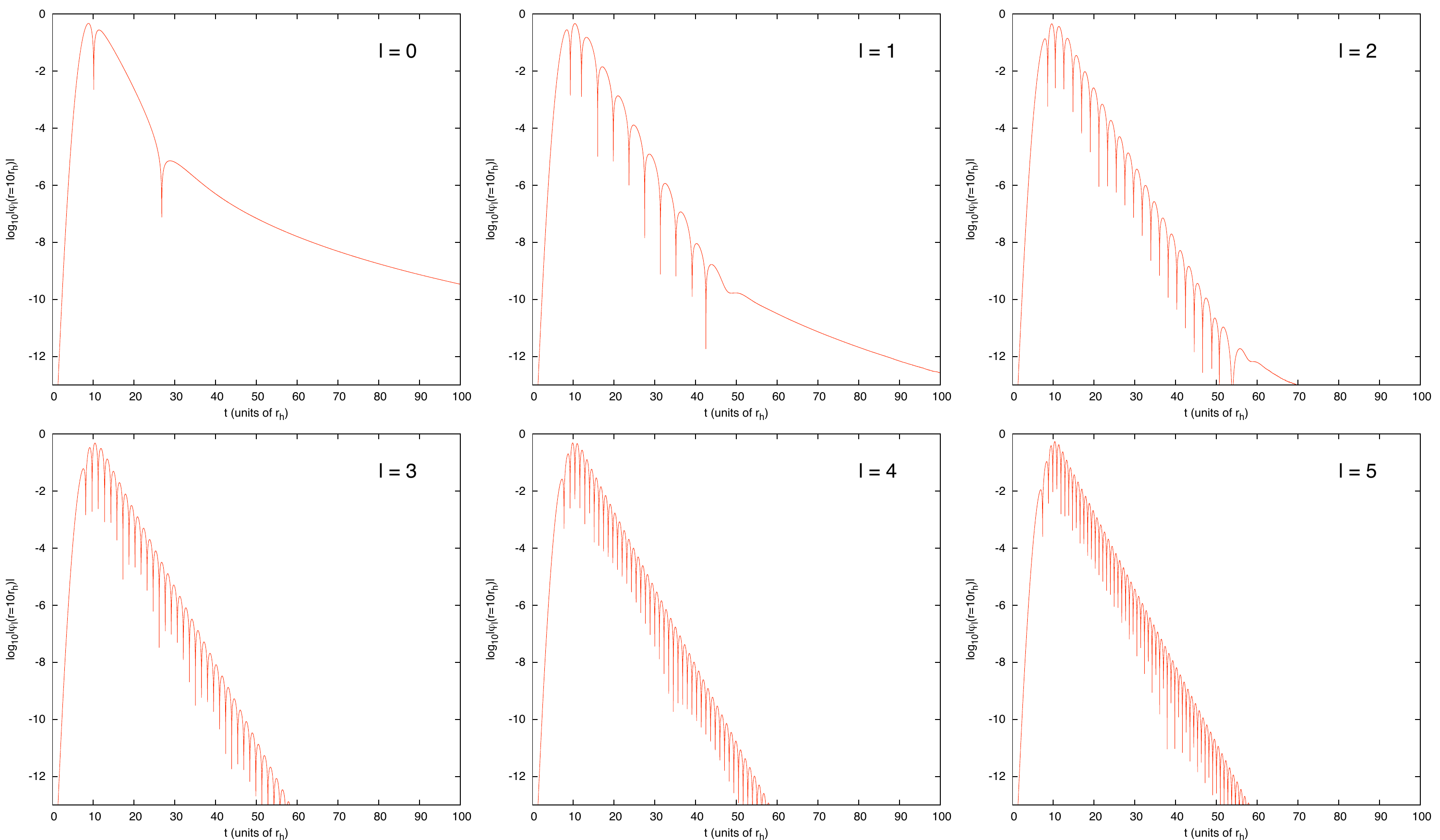}
\end{center}
\caption{Field modes $\psi_l(t,r)$ as a function of time extracted at $r= 10r_h$, for Gaussian initial data $\bar{r}_\ast = 0$, $\sigma = 1$ (see text), for multipoles $l = 0,1,2,3,4,5$. Note logarithmic scale on vertical axis.}
\label{fig-timedomain}
\end{figure*}

Figure \ref{fig-timedomain} shows the field $\psi_l$ extracted at $r = 10 r_h$ as a function of time, for multipoles $l = 0, \ldots, 5$. Modes $l = 1, \ldots 5$ show clear evidence of QN mode ringing, i.e.~regular oscillations with exponential decay. Modes $l=0, 1, 2$ also exhibit late-time power-law decay. 

The frequency of fundamental $n=0$ (i.e.~least-damped) QN modes may be estimated from the time-domain results. The real part is given by the oscillation frequency, and the imaginary part by the exponent of decay. Our estimates are given in the second column of Table \ref{table1}.

\begin{table}
\caption{QN frequencies of the fundamental mode ($n=0$) for $l=1\ldots6$. The second column gives the frequencies extracted from the time-domain simulations of Sec.~\ref{sec-timedomain}. The third column gives the frequency estimate from the expansion method, Eq.~(\ref{n0-freq}). The fourth column lists 6th order WKB results, reproduced from Table III in \cite{Berti-Cardoso-Lemos}. The numeral in parantheses indicates the absolute error in the last displayed digit. The error in the expansion method was estimated from the magnitude of the final terms in the series, Eq.~(\ref{n0-freq}). }
\begin{center}
\begin{tabular}{l | l | l l | l}
\hline \hline
$l$ \; &\;Time Evolution & \multicolumn{2}{|l|}{\;Expansion Method} &  \; WKB   \\
\hline
$1$ & \, $0.818(7) \phantom{5} -  0.608(3) i $  & \, $0.9(2)$ & $ - \, 0.5(1) i$ & \, $1.09 - 0.39 i$ \\
$2$ & \, $1.478(2) \phantom{5} - 0.618(2) i$  & \, $1.49(1)$ & $ - \, 0.61(1) i$ & \, $1.41 - 0.70 i$ \\
$3$ & \, $2.1202(2) - 0.6194(2) i$  & \, $2.121(3)$ & $ - \, 0.618(4) i$ & \, $2.12 - 0.62 i$ \\
$4$ & \, $2.7521(1) - 0.6201(1) i$  & \, $2.7522(8)$ & $ - \, 0.620(1) i$ & \, $2.75 - 0.62 i$ \\
$5$ & \, $3.3799(1) - 0.6203(1) i$ & \, $3.3799(3)$ & $ - \, 0.6202(6) i$ & \\
$6$ & \, $4.0054(1) - 0.6204(1) i$ & \, $4.0053(1)$ & $ - \, 0.6204(3) i$ & \\
\hline
\hline
\end{tabular}
\label{table1}
\end{center} 
\end{table}

\section{Asymptotic Expansion of Quasinormal Modes\label{sec-QNexpansion}}
Figure \ref{fig-timedomain} shows that QN modes play a role in the evolution of a generic perturbation of a CAH. A range of methods have been developed for finding QN frequencies directly from the ordinary differential equation, Eq.~(\ref{cah-rad-eq}), rather than from evolving the partial differential equation, Eq.~(\ref{cah-pde}); see for example \cite{Berti-2004} for a partial list. For the Schwarzschild black hole, the most accurate numerical method was introduced by Leaver \cite{Leaver-1985}. 

Leaver's method relies on obtaining a three-term recurrence relation from a standard ansatz.  
Unfortunately, we cannot obtain a three-term relation in the case of the CAH. For a non-rotating draining bathtub flow in 2D with radial function $f = 1 - r_h^2 / r^2$ it was shown in \cite{Cardoso-Lemos-Yoshida} that the recurrence relation has four terms. For the CAH with $f = 1 - r_h^4 / r^4$ [Eq.~(\ref{eq-f})] the recurrence relation has six terms. Hence, precisely determining the QN spectrum of the CAH from the ordinary differential equation (\ref{cah-rad-eq}) remains a difficult problem. A direct approach via numerical  integration of (\ref{cah-rad-eq}) is unwise, because the radial solutions $\varphi_{l n}(r)$ diverge towards the horizon and at spatial infinity. Instead, the best results in the literature come from the application of the WKB method, taken to sixth order \cite{Berti-Cardoso-Lemos}. The WKB results are shown in the fourth column of Table \ref{table1}.  

In this section we apply an alternative method, recently introduced in \cite{Dolan-Ottewill}. It provides a good approximation in the regime $l \gg n$, but fails at low multipoles. A chief motivation for the method is the observation that QN mode ringing arises near the peak of the potential barrier, and the peak is associated with the existence of an unstable circular orbit (Sec.~\ref{sec-geodesics}). The aim is to shed light on the connection between QN frequencies and the properties of the unstable orbit detailed in Sec.~\ref{sec-geodesics}.

\subsection{Method\label{sec-expansion-method}}
The key step in the method is the introduction of an ansatz of the form
\beq
\varphi_{l n}(r) = \exp \left( i \omega \int^r \alp(r) d r_\ast \right) \RR(r)  \label{cah-ansatz} .
\eeq
Influenced by the factorized form of the orbital equation (\ref{orbit-eq-1c}), we choose $\alpha$ to be
\beq
\alp(r) = \left( 1 - \frac{r_c^2}{r^2} \right) \left(1 + \frac{r_c^2}{2 r^2} \right)^{1/2} ,
\eeq
where $r_c$ is the circular orbit radius given in Eq.~(\ref{circ-orb-params}). It follows that $1 - \alp^2 =  b_c^2 f(r) / r^2$. Note that $\alp$ changes sign at $r=r_c$. It is straightforward to verify that Eq.~(\ref{cah-ansatz}) satisfies the QN mode boundary conditions (\ref{cah-bc}) if $\RR(r)$ tends to constant limits as $r_\ast \rightarrow \pm \infty$. Substituting (\ref{cah-ansatz}) into radial equation (\ref{cah-rad-eq}) and dividing through by $f(r)$ leads to
\begin{eqnarray}
\left( f \RR^\prime \right)^\prime + 2 i \omega \alpha \RR^\prime \nonumber \\  + \left[ \frac{\omega^2 b_c^2 - L^2 + 1/4}{r^2} + 
i \omega \alpha^\prime - \frac{f^\prime}{r} \right]  \RR &=& 0 ,  \label{mod-rad-eq}
\end{eqnarray}
where ${}^\prime$ denotes differentiation with respect to $r$, and 
\beq
i \alp^\prime =  \frac{i b_c^2}{r^3} \left( 1 + \frac{r_c^2}{r^2} \right) \left( 1 + \frac{r_c^2}{2r^2} \right)^{-1/2}  .
\eeq
To seek the fundamental mode ($n=0$), we proceed by expanding the frequency and wavefunction in powers of $L = l+1/2$, 
\begin{eqnarray}
b_c \omega_{l0} &=&  \varpi_{-1} L + \varpi_{0} + \varpi_{1} L^{-1} + \ldots   \nonumber \\
\RR &=& \exp \left( S_{0} + L^{-1} S_{1} +  L^{-2} S_{2}  + \ldots   \right)   \label{S-expansion}
\end{eqnarray}
where $S_k = S_k(r)$ are radial functions to be determined. 
Next we substitute (\ref{S-expansion}) into (\ref{mod-rad-eq}) and collect like powers of $L$ to obtain the system of equations, 
\begin{eqnarray} 
\left( \varpi_{-1} \right)^2 - 1=0, \\
2 i \alp b_c^{-1} \varpi_{-1} S_0^\prime + \frac{2 \varpi_{-1} \varpi_{0} }{r^2}
+ i \alp^{\prime} b_c^{-1} \varpi_{-1} =0,
\end{eqnarray}
\vspace{-1.0cm}
\begin{eqnarray}
\left( f S_0^\prime \right)^\prime +f \left( S_0^\prime \right)^2 
+ 2 i \alpha b_c^{-1} \left( \varpi_{-1} S_1^\prime + \varpi_0 S_0^\prime   \right) &&  \\
+\frac{(\varpi_{0})^2 + 2 \varpi_{-1} \varpi_{1} + 1/4}{r^2} + \frac{i \varpi_0 \alp^\prime }{b_c} - \frac{f^\prime}{r} &&=  0 
, \nonumber
\end{eqnarray}
etc. Now we impose a continuity condition upon $S_k(r)$ at $r= r_c$ to solve the equations for the unknowns $\omega_{k}$ and $S_k^\prime(r)$, as detailed in \cite{Dolan-Ottewill}. 

\subsection{Frequency Expansions}
We obtain for the fundamental ($n=0$) mode
\begin{eqnarray}
\hspace{-0.3cm} b_c \omega_{l0} &=& L - i   - \frac{61}{216L} - i \frac{17}{972L^2} - \frac{532843}{2519424L^3} \nonumber \\
&&  + i \frac{4802843}{5668704L^4} + \frac{11506101785}{4897760256L^5} + \ldots \label{n0-freq}
\end{eqnarray}

Numerical values obtained from frequency expansion (\ref{n0-freq}) are given in Table \ref{table1}, where they are compared against time domain and sixth-order WKB results \cite{Berti-Cardoso-Lemos}. 

Following \cite{Dolan-Ottewill}, we may seek higher overtones with the ansatz 
\beq
b_c \omega_{ln} = \sum^{\infty}_{q=-1} {L^{-q} \varpi_q^{(n)}},
\label{wn}
\eeq
where
\begin{eqnarray}
\chi = \left[
\xi^n +\sum_{i}^{n} \sum_{j}^{\infty} a_{ij}^{(n)} L^{-j} \xi^{n-i}
\right] 
 \prod^{\infty}_{q=0}{ \exp \left( L^{-q} S^{(n)}_q \right)},
\label{Y} \nonumber
\end{eqnarray}
with $\xi \equiv 1- r_c / r $.
 
We find the frequencies of the higher overtones to be
\begin{eqnarray}
b_c \omega_{l n}  \hspace{-0.3cm}&&= L - 2 i N - \frac{240N^2+1}{216 \, L} - \frac{i N (110 N^2 - 19)}{243 \, L^2}  \nonumber \\
&& - \frac{890880N^4 - 375456N^2 + 571027}{2519424 \, L^3}  + \ldots \label{n>0-freq}
\end{eqnarray}
where $N = n+1/2$. Numerical estimates of the QN frequencies for the lowest overtones are given in Table \ref{table3}.

\begin{table}
\caption{Estimates of QN frequencies of low overtones ($n=0$, $1$, $2$) using series expansions, Eq.~(\ref{n0-freq}) [$n=0$] and Eq.~(\ref{n>0-freq}) [$n=1,2$]. Numerals in parantheses indicate the absolute error in the last displayed digit, estimated from the magnitude of the final terms in the series.}
\begin{center}
\begin{tabular}{l | l | l | l}
\hline \hline
$l$ \; & $n=0$ & $n=1$ & $n=2$ \\
\hline
$3$ & $2.121(3)\phantom{5}-0.618(4) i$ & $1.70(2) \phantom{5} - 1.93(7) i$ &  \\
$4$ & $2.7522(8)-0.620(1) i$ & $2.44(1) \phantom{5} - 1.90(4) i$ & $1.74(9) - 3.3(2)i$ \\
$5$ & $3.3799(3)-0.6202(6) i$ & $3.123(6) - 1.89(3) i$ & $2.58(5) - 3.2(1)i$ \\
$6$ & $4.0053(1)-0.6204(3) i$ & $3.790(4) - 1.88(2) i$ & $3.34(3) - 3.2(1)i$ \\
\hline
\hline
\end{tabular}
\label{table3}
\end{center} 
\end{table}

\subsection{Geometric Interpretation}
In the asymptotic regime $l \gg n$, the QN frequencies are
\beq
\omega_{l n} = \Omega \, (l+1/2) - i \Lambda (n+1/2) + \ldots
\eeq
where $\Omega$ is the orbital frequency given in Eq.~(\ref{orbital-freq}) and $\Lambda$ is the Lyapunov exponent given in Eq.~(\ref{lyapunov-exp}). This result was shown for any spherically-symmetric asymptotically flat spacetime in \cite{Lyapunov1}. 

\subsection{Wave functions}
\begin{figure}
\begin{center}
\includegraphics[width=8cm]{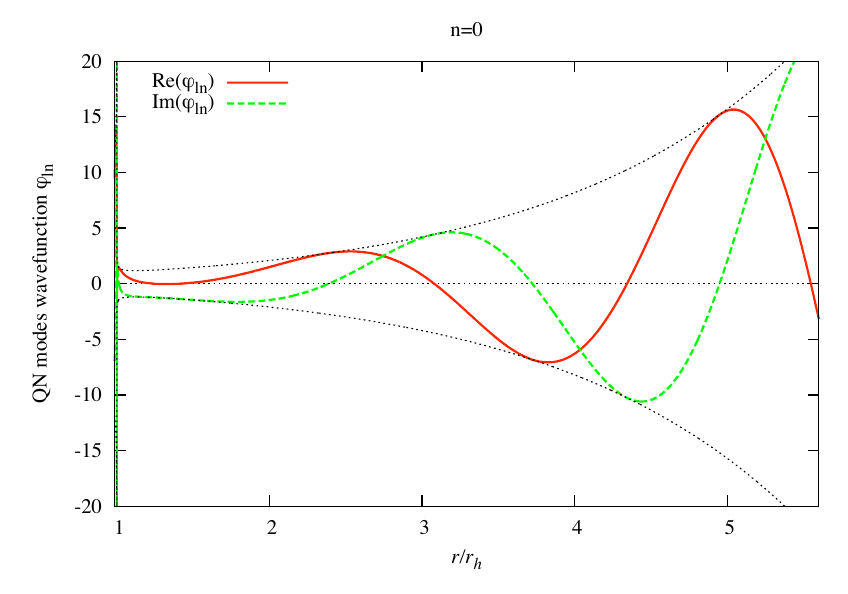} \\
\includegraphics[width=8cm]{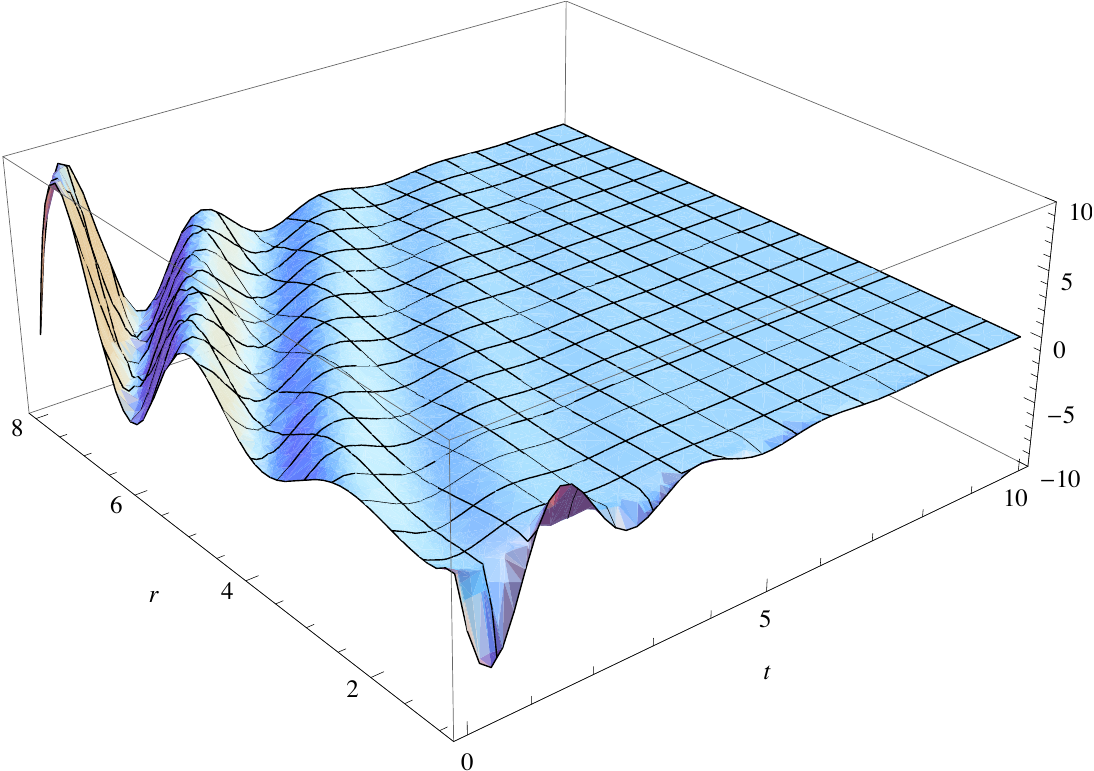}
\end{center}
\caption{\emph{Wavefunction of the fundamental QN mode}. The top plot shows the radial function $\varphi_{l n}(r)$ for $n=0$ and $l=4$. The bottom plot shows $e^{-i \omega t} \varphi_{l n}(r)$ in the ($r,t$) plane. }
\label{fig-QNwavefunction}
\end{figure}

The matching procedure also yields the derivatives of the functions $S_k(r)$, for example
\beq
S_0^\prime = \frac{b_c}{r^2 \alp} - \frac{\alp^\prime}{2 \alp}. 
\eeq
After integration, substitution of $S_k$ into (\ref{S-expansion}) and (\ref{cah-ansatz}) leads to an expansion for the radial wavefunction. A typical example of the radial wavefunction is shown in Fig.~\ref{fig-QNwavefunction}. Note that the wavefunction diverges towards infinity and (more weakly) at the horizon, and decays exponentially with time. This behavior is shown clearly in the lower plot of Fig.~\ref{fig-QNwavefunction}, which depicts the QN mode in the $(r,t)$ plane. It is clear that outgoing ripples are (approximately) constant in the directions $dr =dt$, and there are also (much smaller) ingoing ripples which perturb the sonic horizon. 

\section{Regge Poles\label{sec-RPexpansion}}
It was shown in \cite{Dolan-Ottewill} that the expansion method may also be used to find Regge poles (defined in Sec.~\ref{sec-QNRP}). One approach is to repeat the arguments of Sec.~\ref{sec-expansion-method}. Alternatively, we may simply assume that the Regge poles have the expansion
\beq
\lambda_{\omega n} = L = \zeta_{-1} b_c \omega + \zeta_0 + \zeta_{1} (b_c \omega)^{-1} + \ldots ,
\eeq
then substitute the expansion for QN frequencies into the above equation, and solve order-by-order in $L$ to find the expansion coefficients $\zeta_{k}$. For the fundamental mode we use Eq.~(\ref{n0-freq}) to obtain 
\begin{eqnarray}
\lambda_{\omega, n=0} &=& b_c \omega + i  +  \frac{61}{216 b_c \omega} - \frac{515i}{1944(b_c \omega)^2} 
\nonumber \\ && 
- \frac{291467}{2519424 (b_c \omega)^3} 
- \frac{23295095i}{22674816(b_c \omega)^4}
\nonumber \\ && 
-  \frac{32172386633}{4897760256(b_c \omega)^5} + \ldots  \label{n0-rp}
\end{eqnarray}
For the higher overtones $n$ we use Eq.~(\ref{n>0-freq}) to show
\begin{eqnarray}
\lambda_{\omega n} &=& b_c \omega + 2 i N +  \frac{240N^2 + 1}{216 b_c \omega} - \frac{5i N (344N^2 + 17)}{972 (b_c \omega)^2} \nonumber \\ && - \frac{\left( 8855040N^4 + 1236000N^2 - 570973 \right)}{2519424 (b_c \omega)^3} \nonumber \\ 
&& + \ldots  \label{n>0-rp}
\end{eqnarray}

\subsection{Numerical Method\label{subsec:nummeth}}
Whereas finding QN frequencies via direct numerical integration of (\ref{cah-rad-eq}) is difficult, because the negative imaginary part of the QN frequency leads to divergent behavior in the limits $r_\ast \rightarrow \pm \infty$ (see, e.g., Fig.~\ref{fig-QNwavefunction}), finding Regge poles for real frequencies is much easier, as the solutions are oscillatory in both limits. We used a simple direct numerical integration scheme to test the validity of the expansions (\ref{n0-rp}) and (\ref{n>0-rp}).

Our numerical scheme is based on the shooting method and numerical minimization, and is straightforward to implement. First we choose a complex value of $L = l+1/2$ and construct: (i) the ingoing solution $\psi_{\text{hor}}$ at the horizon written as a generalized power series in $r - r_h$, and (ii) the outgoing solution $\psi_{\text{inf}}$ at infinity written as a generalized power series in $1/r$. Then we use (i) as an initial condition close to the horizon (e.g. $r - r_h = 10^{-3} r_h$), and integrate (\ref{cah-rad-eq}) outwards, and use (ii) as an initial condition far from the horizon (e.g. $r \approx 30 r_h$) and integrate (\ref{cah-rad-eq}) inwards. At some intermediate radius (typically $r_{m} = 5$) we choose the normalisation such that $\psi_{\text{hor}}(r_m) = \psi_{\text{inf}}(r_m)$ and construct the quantity
\beq
\Delta( L ) = \frac{ \psi^\prime_{\text{hor}}(r_m) - \psi^\prime_{\text{inf}}(r_m) }{ \psi^\prime_{\text{hor}}(r_m) + \psi^\prime_{\text{inf}}(r_m) }
\eeq
where ${}^\prime$ denotes the radial derivative. Regge pole values $\lambda_{\omega n}$ correspond to zeros of this function, $\Delta(L = \lambda_{\omega n}) = 0$.  
We employed a numerical root finder to locate the zeros of $\Delta$ in the complex-$L$ plane. 

\subsection{Results}
\begin{figure}
\begin{center}
\includegraphics[width=8cm]{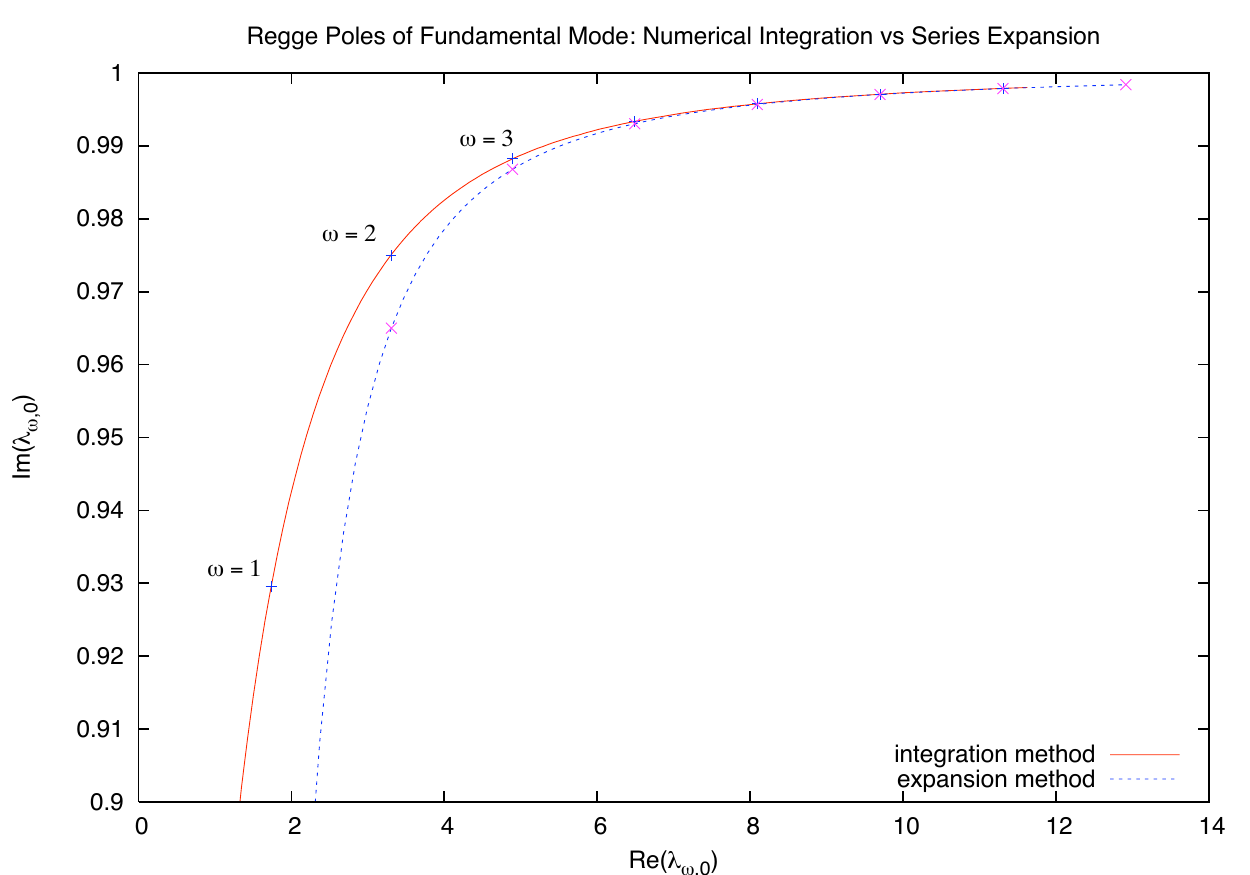}
\end{center}
\caption{Real and imaginary parts of fundamental Regge pole $\lambda_{\omega, n=0}$, shown in frequency range $0.8 < \omega < 8$. The solid [red] line shows the results of the numerical method of Sec.~\ref{subsec:nummeth}. The dotted [blue] line shows the results of the expansion method, Eq.~(\ref{n0-rp}).}
\label{fig-reggepole}
\end{figure}

Regge pole values for the fundamental mode are listed in Table \ref{table-rp}. The results of the expansion method are compared against the numerical integration method. We find excellent agreement at high frequency, but the expansion fails to give an accurate estimate at low frequency, as expected. This behavior is illustrated in Fig.~\ref{fig-reggepole} which shows the real and imaginary part of $\lambda_{\omega, n=0}$ as a function of $\omega$.

\begin{table}
\caption{Regge poles $\lambda_{\omega n}$ of the fundamental mode ($n = 0$) for a range of frequencies $0.25 \le \omega \le 6$. The second column shows the results of the numerical method of Sec.~\ref{subsec:nummeth}. The final column gives estimates from the asymptotic series, Eq.~(\ref{n0-rp}). Note that the final term in Eq.~(\ref{n0-rp}) has not been used due its detrimental effect on accuracy at frequencies $\omega \lesssim 4$. In parantheses we indicate the absolute error in the last displayed digit, estimated from the magnitude of the final terms used in Eq.~(\ref{n0-rp}).}
\begin{center}
\begin{tabular}{l | l | l }
\hline \hline
$\omega$ \; & \; Num.~method & \; Expansion Method \\
\hline
$0.25$ & \; $0.5737 + 0.7708 i$ &  \\
$0.5$  & \; $0.9622 + 0.8569 i$ &  \\
$1.0$  & \; $1.7344 + 0.9295 i$ & \; $1.76(3) \phantom{55} + 0.7(2) i$ \\
$2.0$  & \; $3.3026 + 0.9750 i$ & \; $3.308(3) \phantom{5} + 0.965(9) i$ \\
$3.0$  & \; $4.8916 + 0.9882 i$ & \; $4.893(1) \phantom{5} + 0.987(2) i$ \\
$4.0$  & \; $6.4904 + 0.9933 i$ & \; $6.4908(4) + 0.9930(6) i$ \\
$5.0$  & \; $8.0939 + 0.9958 i$ & \; $8.0941(2) + 0.9957(2) i$ \\
$6.0$  & \; $9.7001 + 0.9971 i$ & \; $9.7002(1) + 0.9971(1) i$ \\
\hline
\hline
\end{tabular}
\label{table-rp}
\end{center} 
\end{table}

\section{Discussion and Conclusion\label{sec-conclusion}}
In this paper we have applied three methods to find the quasinormal modes and Regge poles of the canonical acoustic hole. Let us comment first on the strengths and weaknesses of the methods used, and then on the physical implications of our results.

We showed that extracting QN frequencies from the results of a time domain simulation \cite{Press, Dorband} is an effective way to obtain the fundamental QN mode (for $l > 0$) which clearly dominates the QN ringing phase (Fig.~\ref{fig-timedomain}). However, the method is not suited to finding higher overtones, which decay more rapidly. Time domain simulations of the CAH were previously attempted in \cite{Xi-Li}, but the results presented there do not seem correct. Time domain simulations of acoustic black holes in Laval nozzles may be found in \cite{Okuzumi-Sakagami}.

The expansion method \cite{Dolan-Ottewill} has the advantage of revealing the connection between the properties of the null circular orbit of the effective spacetime, and the large-$l$ limit of the QN frequency spectrum. We used the method to obtain QN frequencies as a series in inverse powers of $l+1/2$ (see Eq. (\ref{n0-freq}) and Eq. (\ref{n>0-freq})). Our series give good estimates in the regime $l \gg n$, but not for low multipoles $l \lesssim n$. Our series have a similar range of validity to the results of the WKB method \cite{Berti-Cardoso-Lemos}. It turns out to be much more computationally-intensive to obtain high-order terms in the frequency expansion for the CAH than for the Schwarzschild black hole. This is because of the difficulty in obtaining a series expansion around $r=r_c$ due to the power $r^{-4}$ in the radial function (\ref{eq-f}). The higher power in $f(r)$ is also the reason why Leaver's method  \cite{Leaver-1985} may not be applied without modification (see Sec.~\ref{sec-QNexpansion}) to the CAH case.

An advantage of the expansion method is that it leads us directly to the asymptotic Regge pole spectrum [Eq.~(\ref{n0-rp}) and Eq.~(\ref{n>0-rp})]. To check the Regge pole values obtained via the expansion method, we used a direct numerical integration scheme (Sec.~\ref{sec-RPexpansion}) to locate Regge poles precisely. The numerical method has the benefit of being exact, up to desired accuracy. Unfortunately the method cannot be reliably used to find QN modes, because of the asymptotic behavior of the radial functions.

Now let us comment on the implications of our results. We have shown (Fig.~\ref{fig-timedomain}) that quasinormal mode ringing is a key feature of the response of a CAH to a generic weak perturbation, and have obtained the physically-relevant part of the QN spectrum. Given the importance of QN ringing in, for example, the search for black hole mergers, we consider the experimental verification of a QN ringing phase in an analogue system to be an important objective. We hope that accurate estimates of frequencies and decay times for QN modes presented here will motivate an experimental study. However, we should not underestimate the experimental challenges in this regard. The CAH is formed from a radially-ingoing fluid flow which becomes supersonic at the `horizon' radius. Typically, at the point of transition between sub- and supersonic flow, a fluid flow becomes unstable, leading to the development of shock waves. This is related to the fact that in the supersonic regime, an accelerated flow diverges rather than converges. Various authors have proposed the use of Laval nozzles to achieve a stable transition between sub- and supersonic regimes. Some possible setups are considered in \cite{Abdalla-Konoplya, Okuzumi-Sakagami, Furuhashi-Nambu-Saida, Cardoso-2005}. Obviously, the QN spectrum depends on the setup, but we expect that the methods used here can be readily adapted to physically-relevant cases.

The Regge pole spectrum computed here may also be of experimental interest. For example, the angular width of the `glory' diffraction pattern that arises from the scattering of a monochromatic wave \cite{doc} is related to the Regge pole spectrum \cite{Decanini-Folacci-Jensen, Andersson} (though it was shown \cite{doc} that the glory effect is much weaker for a CAH than a similarly-sized Schwarzschild hole). Finally, we hope that the simple methods used here may also be applied to other analogue systems of experimental interest in future. 

\begin{acknowledgments}
The authors would like to thank Conselho Nacional de Desenvolvimento 
Cient\'\i fico e Tecnol\'ogico (CNPq) and Funda\c{c}\~ao de 
Amparo \`a Pesquisa do Estado do Par\'a (FAPESPA)
for partial financial support, 
as well as the Geophysics Graduate Program at the 
Universidade Federal do Par\'a (UFPA), in the person of
C\'\i cero R. T. R\'egis, for the use of their computational facilities. 
S. D. thanks the Universidade Federal do Par\'a (UFPA) 
in Bel\'em for kind hospitality, and acknowledges financial support from the Engineering and Physical
Sciences Research Council (EPSRC) under grant no. EP/G049092/1.
L. C. and L. O. would like to acknowledge also partial financial 
support from Coordena\c{c}\~ao de Aperfei\c{c}oamento de Pessoal
de N\'\i vel Superior (CAPES).
\end{acknowledgments}

\bibliographystyle{apsrev}


\begin{thebibliography}{20}
\bibitem{time-footnote}
Here $t$ is a non-physical time coordinate related to the laboratory time coordinate $\bar{t}$ via 
$dt = d\bar{t} - (r_h^2/r^2) (1-r_h^4/r^4)^{-1} dr$.
\end{thebibliography}

\end{document}